\documentclass{aa}
\usepackage{graphicx}
\usepackage{lscape}
\usepackage{natbib}
\bibpunct[ ]{(}{)}{;}{a}{}{,}

\def\spose#1{\hbox to 0pt{#1\hss}}
\def\lta{\mathrel{\spose{\lower 3pt\hbox{$\mathchar"218$}}
        \raise 2.0pt\hbox{$\mathchar"13C$}}}
\def\gta{\mathrel{\spose{\lower 3pt\hbox{$\mathchar"218$}}
        \raise 2.0pt\hbox{$\mathchar"13E$}}}

\begin{document}

 \title{Tidal torques, disc radius variations, and instabilities in dwarf novae and soft X-ray transients}
 \authorrunning{Hameury and Lasota}
 \titlerunning{Disc radius variations in dwarf novae}
 \author{Jean-Marie Hameury\inst{1} \and Jean-Pierre Lasota\inst{2}}
 \institute{ UMR 7550 du CNRS, Observatoire de Strasbourg, 11 rue
             de l'Universit\'e, F-67000 Strasbourg, France \\
             \email{hameury@astro.u-strasbg.fr}
 \and
             Institut d'Astrophysique de Paris, UMR 7095 CNRS, Universit\'e Pierre \&
Marie Curie, 98bis Boulevard Arago,
             75014 Paris, France\\
             \email{lasota@iap.fr} }
 \offprints{J.M. Hameury}

\date{Received 24 June 2005 / Accepted 29 July 2005}

\abstract{The study of outer disc radius variations in close
binary systems is important for understanding the structure and
evolution of accretion discs. These variations are predicted by
models of both quasi-steady and time-dependent discs, and these
predictions can be confronted with observations. We consider
theoretical and observational consequences of such variations in
cataclysmic variables and low-mass X-ray binaries. We find that
the action of tidal torques, that determine the outer radius at
which the disc is truncated, must be important also well inside
the tidal radius. We conclude that it is doubtful that the
tidal-thermal instability is responsible for the
superoutburst/superhump phenomena in dwarf novae, and confirm that
it cannot be the reason for the outbursts of soft X-ray
transients. It is likely that tidal torques play a role during
superoutbursts of very-low mass-ratio systems but they cannot be
the main and only cause of superhumps.
 }

\maketitle

\section{Introduction}

Dwarf novae are a subclass of cataclysmic variables that undergo
outbursts lasting at least a few days during which their
brightness increases by several magnitudes \citep[see e.g.][for a
review]{w95}. These outbursts are believed to be due to a
thermal/viscous accretion disc instability which arises when the
disc temperature becomes of order of $\lta$10$^4$K, enough for
hydrogen to become partially ionized and opacities to depend
strongly on temperature \citep[see][for a review of the
model]{l01}. Models have become more and more sophisticated, and
differ from the initial, basic version of the model proposed by
\citet{mm81} and by \citet{s82}, following precursor works by
\citet{s71}, \citet{o74} and \citet{h79}. They now include various
effects such as illumination of the disc by the white dwarf
\citep{hld99}, disc truncation \citep{hlmn97} by evaporation or by
a magnetic field, increase of the mass transfer rate from the
secondary as a result of illumination \citep{hlw00}, and heating
of the outer parts of the disc by the stream impact and tidal
torques \citep{bhl01}. The weakest point of these models -- aside
from the assumption that angular momentum transport is due to a
``viscosity" (i.e. is a local phenomenon accompanied by energy
dissipation) described by the modified alpha prescription of
\citet{ss73} (bivalued $\alpha$, low in quiescence and high in
outburst), is the approximate treatment of intrinsically 2D
effects at the disc edge.

There is in particular a debate about the outcome of the disc
reaching the radius at which the 3:1 resonance occurs (this may
happen for low secondary to primary mass ratios). SPH models
\citep[see e.g.][]{w98,m96,m98} treat accurately the dynamics of
the disc and are in principle quite appropriate to deal with these
effects; they predict that when the 3:1 resonance radius is
reached, the disc becomes tidally unstable, eccentric and
precesses. SPH models are however limited in that it is difficult
to include in them detailed microphysics; the energy equation is
often replaced by isothermal or adiabatic approximations. This
significantly affects the results as shown by \citet{kr00} who
found that the tidal instability does not occur when disc cooling
is properly taken into account but is present when the disc is
assumed to be isothermal. Nevertheless, the tidal instability,
coupled with the standard thermal instability, is also believed
\citep{o89} to be the reason for the long duration of
superoutbursts in SU UMa systems, a subclass of dwarf novae
exhibiting large and long outbursts during which so called
superhumps are seen, together with normal outbursts. These
superhumps are light modulations at a period slightly longer than
the orbital period, and it is often taken for granted that they
result from precession of a distorted disc.

Another point of debate is the amplitude of the torque $T_{\rm
tid}$ due to the tidal forces, even far from the resonance. This
torque transfers the outward-flowing angular momentum from the
accretion disc back to the orbit, effectively truncating the disc,
and therefore determining its radius. Based on their numerical
simulations \citet{io94} concluded that $T_{\rm tid}$ should be
negligible everywhere in the disc, except in a small ring at the
tidal truncation radius; on the other hand, it has been often
assumed that $T_{\rm tid}$ has a smoother behaviour. In this
paper, we first discuss (section 2) the effect of two different
prescriptions for $T_{\rm tid}$ on the light-curve of systems for
which we know that the disc does not reach the 3:1 resonance. We
show that the predicted light curves are not sufficiently
different to be discriminated by observations; e.g. the
determination of outer disc radius in eclipsing dwarf novae is not
accurate enough to clearly rule out one of the models, even though
a smooth $T_{\rm tid}$ appears to be in better agreement with
observations. We compare the predictions of the disc instability
model using these prescriptions for VY SCl stars (section 3),
which exhibit slow variations of their luminosity from a high to a
low state and back to high luminosity. We conclude that $T_{\rm
tid}$ cannot be neglected even in the central parts of the disc.
We then turn to SU UMa systems (section 4) and soft X-ray
transients (SXTs) containing a black hole (section 5), and examine
for these systems the consequences of assuming that superhumps are
due to a tidal instability of the disc. We show that in such a
case, SXTs with short orbital periods should have detectable
superhumps, present both in quiescence and in outburst, but no
superhump is expected during outbursts and quiescence of long
period X-ray binary systems.

\section{Disc radius variations in dwarf novae}

The angular momentum conservation equation for a Keplerian disc can
be written as:
\begin{eqnarray}
j \frac{\partial \Sigma}{\partial t} = - \frac{1}{r}
\frac{\partial}{\partial r} \left(r \Sigma j v_{\rm r}\right) & +
& \frac{1}{r} \frac{\partial}{\partial r}
\left(- \frac{3}{2} r^2 \Sigma \nu \Omega_{\rm K} \right) \nonumber\\
& + & \frac{j_2}{2 \pi r}\frac{\partial \dot{M}_{\rm tr}}{\partial
r} - \frac{1}{2 \pi r} T_{\rm tid}(r) \label{eq1}
\end{eqnarray}
where $\Sigma$ is the surface column density, $\dot{M}_{\rm tr}$ is
the rate at which mass is incorporated into the disc at radius $r$,
$v_{\rm r}$ is the radial velocity in the disc, $j = (G M_1
r)^{1/2}$ is the specific angular momentum of material at radius $r$
in the disc ($M_1$ being the primary mass), $\Omega_{\rm K} = (G
M_1/ r^3)^{1/2}$ is the Keplerian angular velocity, $\nu$ is the
kinematic viscosity coefficient, and $j_2$ the specific angular
momentum of the material transferred from the secondary. $T_{\rm
tid}$ is the torque due to the tidal forces. One often use the
prescription of \citet{s84}, derived from the linear analysis of
\citet{pp77} which, for radii not too close to the tidal truncation
radius $r_{\rm tid}$ at which the trajectories of test particle
orbiting around the white dwarf intersect leading to high
dissipation \citep{p77}, was confirmed by the non-linear numerical
simulations of \citet{io94}. This reads:
\begin{equation}
T_{\rm tid} = c \omega r \nu \Sigma \left(\frac{r}{a}\right)^n
\label{torque}
\end{equation}
where $\omega$ is the angular velocity of the binary orbital
motion, $a$ the binary orbital separation and $c$ a numerical
constant. The index $n$ was found to be close to $n=5$. In the
particular case of \object{U Gem}, \citet{io94} obtained a value
for $c$ smaller by almost two orders of magnitude than the value
required for the disc to be truncated at the tidal radius in
steady state. They concluded that the tidal torque is very small
everywhere except very close to the tidal radius where $T_{\rm
tid}$ diverges. If true, the tidal removal of the angular momentum
would occur only at the disc's outer edge within a negligibly
small radial extent \citep{s02}.

\begin{figure}
 \resizebox{\hsize}{!}{\includegraphics{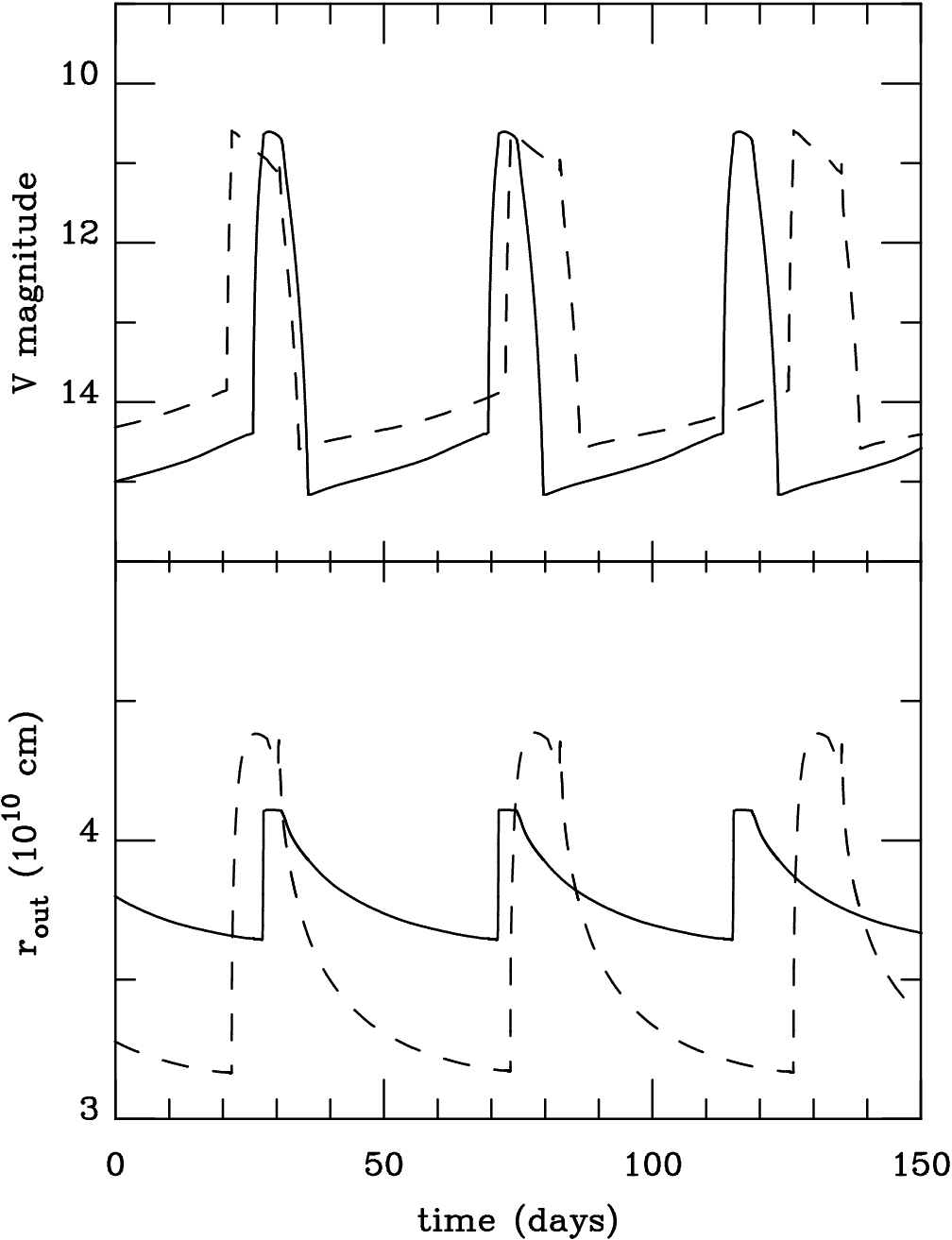}}
 \caption{Outburst
models in which the tidal torque is either vanishingly small for
$r < r_{\rm tid}$ and grows exponentially on a short scale for
larger $r$ (solid curves), or is given by Eq. \ref{torque} (dashed
curve). The top label shows the time variations of the visual
magnitude (arbitrary distance), and the bottom panel gives the
radius of the disc outer edge.}
\end{figure}

As emphasized by \citet{s02}, the tidal torque would then simply
force the disc to remain within the tidal radius; contraction of
the disc would result from accretion of matter with low angular
momentum during quiescent phases. In order to evaluate the effect
of various prescriptions for the tidal torque, we used our
disc-instability model code \citep[][\hspace*{-0pt}; see also
Buat-M{\'e}nard et al., 2001]{hmdl98} to calculate models with two
different prescriptions for the tidal torque: $T_{\rm tid} \propto
\exp((r-r_{\rm tid})/10^8\; \rm cm)$, vanishingly small for $r <
r_{\rm tid}$ and growing exponentially on a short scale ($\ll
r_{tid}$) for larger $r$ (prescription (A)), and $T_{\rm tid}$
given by Eq. \ref{torque}, in which the constant $c$ has been
adjusted to produce a given average radius (prescription (B)). Our
code is a 1D + 1D scheme solving the radial angular-momentum and
energy equations decoupled from the vertical structure equations
whose solutions are represented by a grid of (pseudo)thermal
equilibria \citep[see][for details]{hmdl98}. The viscosity is the
standard \citet{ss73} one; here, we take $\alpha_{\rm cold} =
0.04$ and $\alpha_{\rm cold} = 0.2$, and assume a smooth (but
rapid) transition) between the value of $\alpha$ on the cool and
hot branch \citep[see][for more details]{hmdl98}. We have also
assumed that in case (A), the dissipation of the tidal torque
results in enhanced radiation from the disc edge, as suggested by
\citet{s02}, important only when the outer radius is close to the
tidal radius, i.e. during outbursts, and therefore does not affect
the stability of the disc, contrary to what happens in case (B),
where an additional heating term has to be taken into account
\citep{bhl01}. The parameters of the system are those of \object{U
Gem} (primary mass $M_1$ = 1 M$_\odot$, orbital period $P_{\rm
orb}$ = 4.25 hr, tidal radius 4.06 10$^{10}$, circularization
radius 1.30 10$^{10}$ cm, average mass transfer rate from the
secondary 5 10$^{16}$ g $^{-1}$).

As can be seen, the general shape of the light curve is not
strongly affected by this radical change of the prescription for
the tidal torque. The recurrence time is almost the same for both
prescriptions, the outburst light curve is not much modified,
except that the duration is longer for prescription (B) than for
prescription (A); this is a consequence of the disc heating by the
dissipation of the tidal torque. Such a small difference between
both sets of light curves contrasts with the large differences
appearing when one assumes or not a fixed outer radius
\citep{hmdl98}. There is however a significant difference between
the evolution of the outer disc radius in both cases, as expected:
in case (B), the existence of a significant torque for $r < r_{\rm
tid}$ results in a significant contraction of the disc during
quiescence, and, conversely, the disc can also extend to larger
radii (possibly beyond $r_{\rm tid}$ !) because of the smoother
variation of the tidal torque. The average disc size is identical
in both cases (by construction), and its variations are smaller in
case (A) than in case (B), as expected. Although this difference
is significant (a factor 2.5 in the amplitude of the outer disc
radius variations), it is probably not large enough to be
unambiguously constrained by observations. We note for example
that \citet{s96} found both in the case of U Gem and Z Cha that
the outer disc radius varies by about 30\%, which is quite
compatible with the amplitude for prescription (B) -- 37 \%, but
we would certainly not dismiss prescription (A) on this sole
basis, in view of the large observational and systematic
uncertainties in the measurement of the outer disc radius.

It is also worth to be noted that, if in case (A) radiation by the
disc edge were not sufficient to get rid of all of the energy
dissipated by tidal heating, the overall light curve would show
marked differences from Fig. 1. Because, as a consequence of the
very steep variations of $T_{\rm tid}(r)$, the energy would be
released in a small annulus, the local energy dissipation rate
could then be quite high in the vicinity of the tidal radius; this
would result in the outer ring being quite hotter that the
remaining outer disc part, with a temperature quite sensitive to
radius variations. As a result, $\sim$ 1 magnitude oscillations
lasting for a few days would be produced shortly after the end of
an outburst. Apparently such oscillations have not been observed,
but their existence certainly depends on the details of the model
assumptions, and in particular on the radial heat transfer at the
very disc edge.

\section{VY Scl systems}

VY Scl stars are a subgroup of cataclysmic variables which are
usually in a bright high state, and have occasionally low states
during which their luminosity drops by more than one magnitude,
bringing them into the dwarf nova instability strip. Yet, they do
not have dwarf nova outburst, even though the decline can be very
gradual and prolonged, longer than the disc viscous time. It had
been suggested that the apparent stability of VY Scl discs could
be due to the irradiation of the inner disc by a very hot white
dwarf \citep{lhk99}, but \citet{hl02} showed that outbursts are
unavoidable unless the disc disappears completely during
quiescence, most probably truncated by the white dwarf magnetic
field. The disc has to remain hot at all times so that a cooling
front does not start from the outer edge, until it completely
disappears. This translates into a requirement on the magnetic
field strength that must be sufficient for the Alfv\'en radius to
be equal to the circularization radius when the accretion rate is
just equal to the critical rate below which instabilities appear.

This model has received strong observational support \citep{hl05}.
Observations of \object{MV Lyr} in a low state show no indication
for a disc; \citet{lsg05} have put an upper limit of 2500 K on the
disc temperature. Similarly, \object{TT Ari} also shows
(virtually) no sign for an accretion disc in its low state
\citep{gsb99}.

These systems are therefore a key in understanding how tidal
torques act on a disc, since they slowly vary from a state in
which there is a fully developed disc to a discless state, and the
outer disc radius has to vary continuously from the tidal
truncation radius to the circularization radius.

\begin{figure}
 \resizebox{\hsize}{!}{\includegraphics[angle=-90]{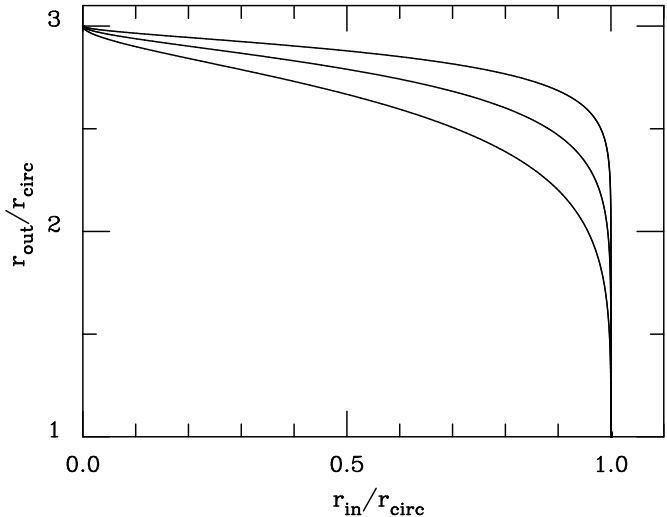}}
 \caption{Variations of the outer radius as a function of the
 inner disc radius, for different values of $n$ in Eq. 5. From top
 to bottom: $n=21$, $n=11$, and $n=6$. The latter corresponds to
 prescription A. The constant $K$ has been chosen such as to give
 the same $r_{\rm out}$ for $r_{\rm in} = 0$.}
 \label{fig:rout}
\end{figure}
Recently, \citet{skb04} observed the eclipsing VY Scl star
\object{DW UMa} in a state intermediate between minimum and
maximum. Eclipse mapping techniques allowed to reconstruct the
disc luminosity profile; they found that the luminosity difference
between the high and intermediate states is almost entirely due to
a change in the accretion disc radius, from $\sim$ 0.5 to $\sim
0.75$ times $R_{\rm L_1}$, the distance between the white dwarf
and the $L_1$ point. \citep[][have caught \object{MV Lyr} in a
similar intermediate state.]{lsg05} It is observed that in the
intermediate state, the disc is entirely eclipsed by the
secondary, while its outer parts are visible during the high
state. A possible explanation of this could be that the disc is
not in equilibrium (i.e. $\dot{M}$ is not constant within the
disc), so that radius variations are a consequence of mass
redistribution inside the disc. This however cannot be the case,
since, as noted by \citet{skb04}, the recovery from the low state
is slow and takes $\sim$ 4 months; if the viscosity is not
unusually small, the disc has enough time to readjust its
structure to changes of the mass accretion rate. The disc should
then be quasi steady, and its radius close to the tidal radius if
tidal torques were exerted only in a very small annulus at $r \sim
r_{\rm tid}$ (prescription A). Observations show, however, a disc
radius $\sim 50\%$ smaller than $r_{\rm tid}$ \citep{skb04}.
Assuming a value of $\alpha$ small enough that the viscous time is
comparable to or longer than the rise-time would of course solve
the problem, but there is absolutely no reason for doing so,
especially that the reason for luminosity variations in VY Scl are
changes in the mass-transfer rate and not modifications of the
disc's physical state. In any case playing with $\alpha$ every
time something does not fit a popular belief is not very helpful.

The relation between the inner and outer radius depends on the
precise prescription for the tidal torque. Assuming steady state,
one can integrate Eq. \ref{eq1} over $r$ and obtain an angular
momentum conservation relation:
\begin{equation}
r_{\rm circ}^{1/2} - r_{\rm in}^{1/2} = {1 \over \sqrt{G M_1}
\dot{M}} \int_{r_{\rm in}}^{r_{\rm out}} T_{\rm tid}(r) dr
\end{equation}
where $r_{\rm circ}$ is the circularization radius (i.e. the
radius at which matter originating from the secondary would
circularize assuming angular momentum conservation), and where we
have used the outer boundary condition \citep{hmdl98}
\begin{equation}
\dot{M} \left[1-\left({r_{\rm circ} \over r_{\rm
out}}\right)^{1/2}\right] = 3 \pi \nu \Sigma
\end{equation}
Writing the tidal torque as
\begin{equation}
T_{\rm tid}(r) = Const. \, \nu \Sigma r^n
\end{equation}
which is valid for both tidal torque prescriptions used here ($n$
arbitrary large for prescription A, and $n=6$ for prescription B);
one finds a solution for the outer radius
\begin{equation}
\int_{u_{\rm in}}^{u_{\rm out}} \exp \left[ {K \over (2 n
+1)}(u^{2n+1} - u_{\rm out}^{2n+1}) \right] du = u_{\rm out} - 1
\end{equation}
where $u_{\rm in} = (r_{\rm in}/r_{\rm circ})^{1/2}$, $u_{\rm out}
= (r_{\rm out}/r_{\rm circ})^{1/2}$, and $K$ is a constant. Figure
\ref{fig:rout} shows the solution of this equation for different
values of $n$. As expected, the disc reduces to a ring becoming
smaller and smaller as $r_{\rm in}$ approaches $r_{\rm circ}$. A
significant and detectable effect is found for $n=6$ (prescription
B), but for larger values on $n$, variations of $r_{\rm out}$ are
small; for example, for $n=11$, a 10\% variation is found when
$r_{\rm in}$ = 0.66 $r_{\rm circ}$, and a 20\% variation requires
$r_{\rm in}$ to be within 7\% of $r_{\rm circ}$. For $n=21$,
$r_{\rm in}$ must be within 11 \% and 0.8 \% respectively of
$r_{\rm circ}$ to get the same variations. The ratio of $T_{\rm
tid}$ at $r_{\rm circ}$ between the three curves is 1:140:450,000;
the $n=11$ curve thus approximately corresponds to the numerical
solution of \citet{io94} in the vicinity of $r_{\rm circ}$, which
therefore appears totally unable to account for the observed
radius variations of \object{DW UMa}.

\begin{figure}
 \resizebox{\hsize}{!}{\includegraphics{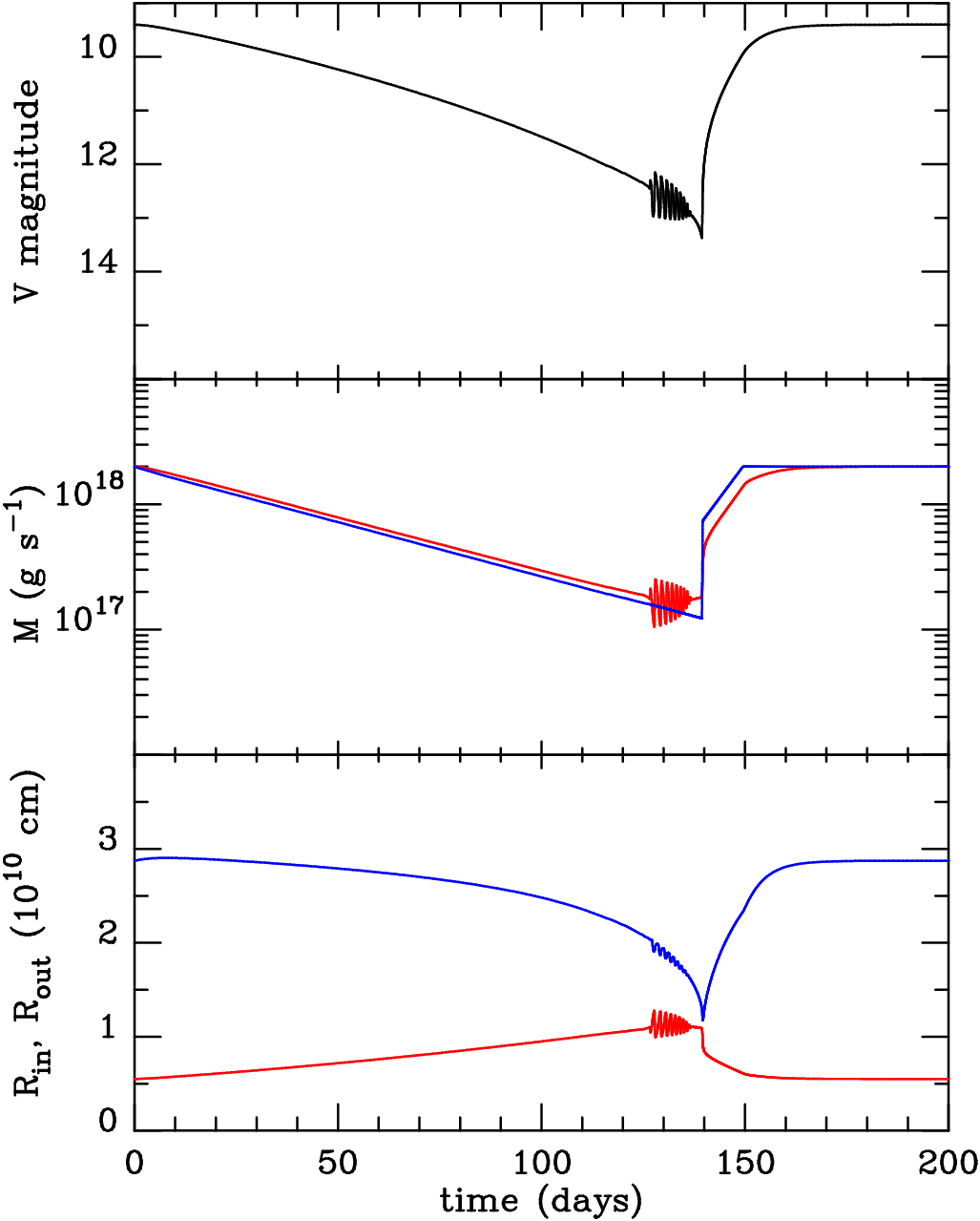}}
 \caption{Evolution of a system with the parameters of \object{DW
 UMa}
 resulting from changes of the mass transfer from the secondary. The
 top panel shows the visual magnitude of the system, the intermediate
 one the mass transfer from the secondary (blue curve) and the mass
 accretion rate onto the white dwarf (red curve, slightly below the
 previous one), and the bottom panel the inner an outer disc radius.
 The magnetic moment is 8 10$^{32}$ G cm$^3$, and the primary mass
 0.7 M$_\odot$ }
 \label{fig:evoldw}
\end{figure}

We used our disc-instability model code \citep[][\hspace*{-0pt};
see also Buat-M{\'e}nard et al., 2001]{hmdl98} to simulate the
accretion-disc properties of \object{DW UMa}. Figure
\ref{fig:evoldw} shows the evolution of a system with parameters
similar to those of DW UMa, in which the white dwarf is magnetized
strongly enough to prevent dwarf nova outbursts during
intermediate states until the disc disappears. We have taken here
$M_1$ = 0.7 M$_\odot$, $M_2$ = 0.3 M$_\odot$, the orbital period
is 3.5 hr, and the outer disc radius is 3 10$^{10}$ cm in the
steady high state; the magnetic moment is 8 10$^{32}$ G cm$^3$,
and we assumed that the disc is illuminated by the hot white dwarf
(temperature of 40,000 K in quiescence). Irradiation by the white
dwarf was treated as in \citet{hl02}. As above, $\alpha_{\rm hot}
= 0.2$ and $\alpha_{\rm cold} = 0.04$. There are still some small
oscillations left but these could be suppressed if the field were
slightly stronger, and are probably not detectable. We have
assumed a slow decrease of the mass transfer rate $\dot{M}_{\rm
tr}$ until the disc almost disappears, followed by a rapid
(essentially for numerical reasons, the adaptative mesh code being
unable to deal with a vanishingly small disc) increase of
$\dot{M}_{\rm tr}$. The second panel shows that the disc is always
close to equilibrium ($\dot M \approx \dot M_{\rm tr}$). It can be
seen that significant variations of the outer disc radius are
obtained. When the disc is fainter by 1 magnitude than the
maximum, the disc is 20\% smaller than its maximal extension. This
is not quite the 50\% variations that are claimed by
\citet{skb04}, but is acceptable in view of the fact that we
define the disc outer edge as the place where the surface density
vanishes, whereas \citet{skb04} use a photometric definition.

These results show that the tidal coupling cannot be negligible in
the intermediate state, even at radii quite a bit smaller than the
tidal truncation radius. It is worth noting in this context that
the mass ratio for this system is $q = 0.39\pm 0.12$, i.e.
marginally compatible with the maximum value of $q$ for the 3:1
resonance to be accessible.

\section{SU UMa systems}

As mentioned earlier, the popular explanation for superoutbursts
in SU UMa stars combines the thermal instability with a tidal
instability that is supposed to arise when the disc reaches the
3:1 resonance radius \citep[see e.g.][]{o89}. At this point, the
disc would become eccentric and precess, allegedly causing the
superhumps. The tidal torque $T_{\rm tid}$ is supposed to increase
by at least one order of magnitude, resulting in a corresponding
enhancement of dissipation and angular momentum transport.
According to this scenario, the disc shrinks until the radius has
decreased to an (arbitrarily chosen) critical value of order of
0.35 times the orbital separation, and the superoutburst stops. A
sequence of several normal outbursts, during which the disc grows
on average then follows until the next superoutburst.

This model is essentially based on SPH simulations indicating that
the disc does become eccentric and precesses when the 3:1
resonance radius is reached \citep{m00,tmw01}, and on the fact
that SU UMa stars are found below the period gap, for systems in
which the mass ratio $q$ is less than 1/3, the condition for the
3:1 resonance radius to be smaller than the tidal truncation
radius. According to simulations by \citet{tmw01} the critical
value of $q$ above which no superoutburst occur is in fact rather
1/4.

However, two elements seem to have seriously put in doubt the
viability of the thermal-tidal model of super- humps and
outbursts. First, \citet{kr00} found that the outcome of numerical
simulations depends on the equation of state that has been used;
they found that eulerian models gave the same results as SPH codes
in the isothermal approximation, but obtained very different
results (no superhumps) when using the full thermal equation in
their eulerian code. More recently, \citet{sw05} found superhumps
in the famous U Gem 1985-superoutburst.The component masses of
this prototypical binary are rather well constrained giving a
rather large value of $q=0.364\pm 0.017$ \citep[see][]{ugem}.
Clearly the tidal instability cannot apply to this system. In
addition, a permanent superhump was detected in \object{TV Col}
\citep{retter03} a binary with an estimated mass-ratio between
$q=0.62$ and 0.93 \citep{hellier93}, consistent with its 5.39 hour
orbital period. One could try to save the tidal model by arguing
that these superhumps are not the usual ones, and that,therefore
\object{U Gem} and \object{TV Col} superhumps are phenomena
different from those observed in SU UMa stars. \citet{sw05}
examined this possibility in detail, and we repeat here their
argument that this is not the case. The superhumps in U Gem have
an amplitude typical of normal superhumps; even more, when plotted
on superhump excess-period vs orbital period (logarithmic) plane,
U Gem and TV Col fall on the linear extension of the relations
defined by shorter period dwarf novae and permanent superhumpers
\citep[see][Fig. 20]{patt03}. Absolutely nothing distinguishes
them from superhumps in low-$q$ systems. One should note in
addition that SPH models fail to reproduce the exact period
excess: this is always found to be larger by a factor 1.5 -- 2
than the measured values, making the position of U Gem on the
superhump excess-period vs orbital period plane even more
significant.

For the moment there are no alternative models for SU UMa's
\citep[but see however][]{b04}. It is likely that irradiation and
enhanced mass transfer play a significant role
\citep{s95,s04,hlw00}, but at present there is no universally
accepted explanation for the superhump/superoutburst phenomenon. It
nevertheless remains true that the tidal torque has to be modified
when the 3:1 resonance is reached; the disc response and the
amplitude of the change are still a matter of debate, but they are
certainly very different that what has been assumed in the TTI
model.

\begin{figure*}
 \begin{center}
 \resizebox{0.4\hsize}{!}{\includegraphics{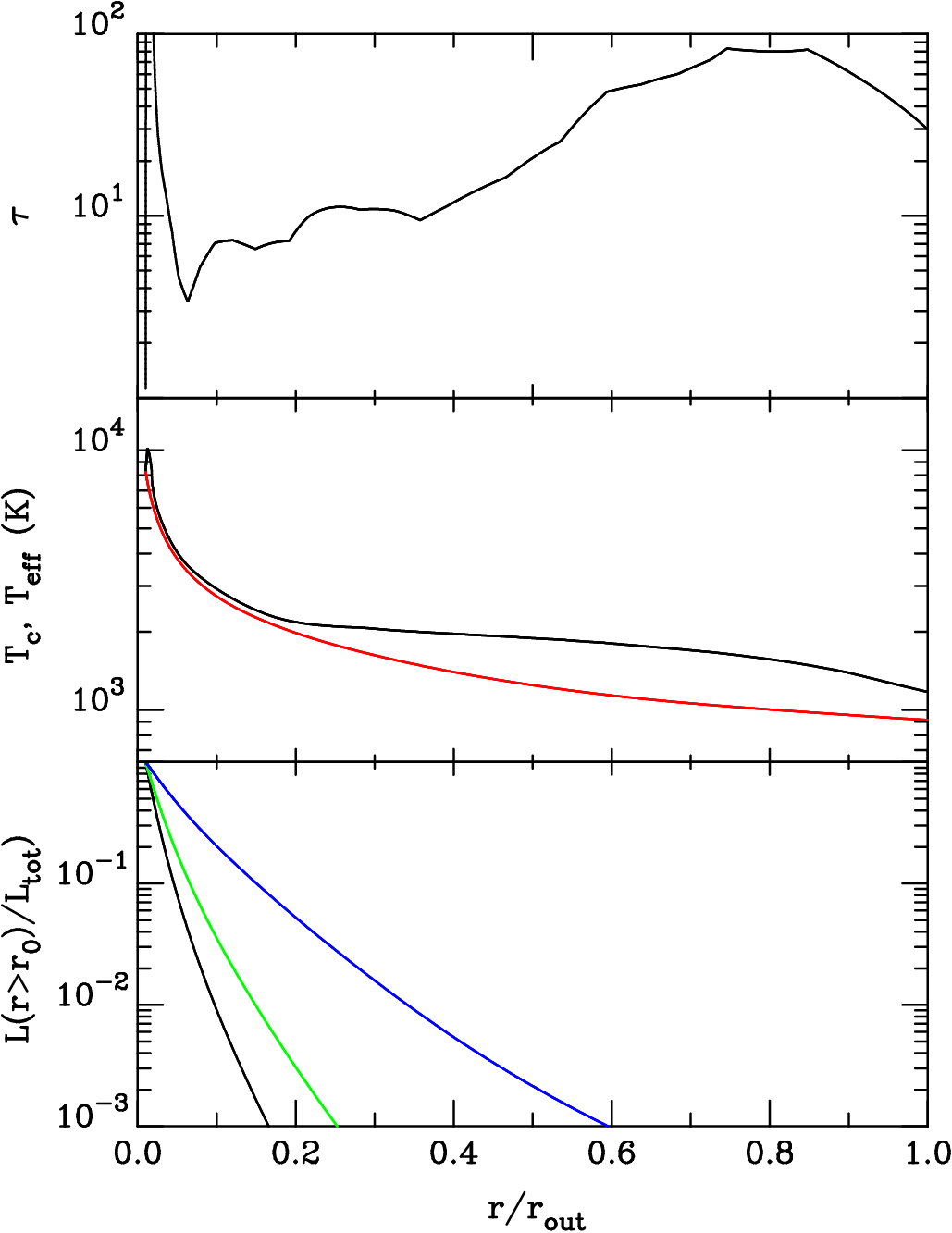}}
 \hspace{0.5cm}
 \resizebox{0.4\hsize}{!}{\includegraphics{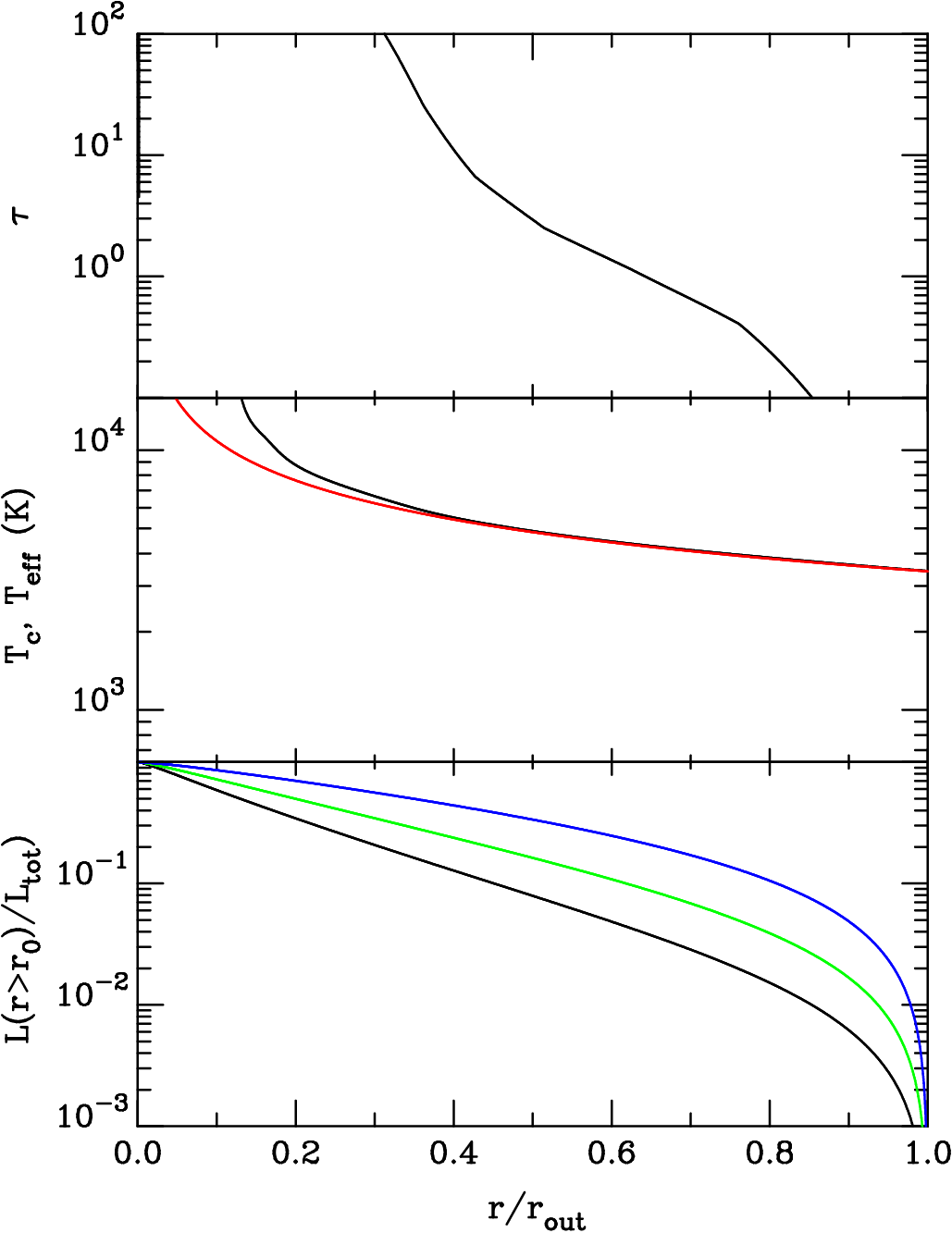}}
 \end{center}
 \caption{Structure of the accretion disc in a long-period system,
 in quiescence (left), and in outburst (right).
 The top panel shows the optical depth of the disc as a function of
 radius, the intermediate one the central (upper curve in black) and effective
 (lower curve, red) temperatures, and the bottom panel gives the fraction
 of the total luminosity emitted above a given radius in the I (blue curve,
 upper), V (green curve, intermediate), and B (black curve, lower)
 bands}
 \label{fig:sxt}
\end{figure*}

\section{Soft X-ray transients}

In soft X-ray transients, the mass ratio is usually very small,
and one would expect to find superhumps in these systems, if
indeed the explanation of this effect is related to the tidal
instability. Indeed there is observational evidence of modulations
at periods slightly longer than the orbital period \citep{z02}.
There are however several important differences between SXTs and
SU UMa systems: first, because the mass ratios are extremely
small, the disc should remain permanently eccentric, and second,
the outer parts of the disc can remain unaffected by the thermal
instability in systems where the disc is large. Finally, in SXTs
superhumps would arise from a modulation of the reprocessed flux
by the changing disc area and not from an increase of viscous
dissipation as in SU UMa stars \citep{hkmc01}. We do not discuss
here problems that might arise from the amount of reprocessing
required to account for the observed modulations, and that have
been addressed by \citet{hkmc01}.

As an example, for $q = 0.05$, the circularization radius is $0.42
a$, where $a$ is the orbital separation, larger than the critical
radius below which the tidal instability cannot be maintained
\citep[typically $0.35 a$, see e.g.][]{o89}. The disc can
therefore never shrink  enough for the tidal instability to stop,
and it should remain in a permanently eccentric state. This will
occur for values of $q$ below 0.10, i.e. for most SXTs. Therefore
the tidal instability cannot be the cause of SXT outbursts -- and,
as far as we know, this has never been suggested. It is then
interesting to note that SXTs have sometimes been compared to WZ
Sge systems \citep[e.g.][]{k00}, a particular class of SU UMa
stars that have no normal outbursts between large and unfrequent
superoutbursts. If the observational similarities between both
classes reflect similar physics, this is then a additional problem
for the tidal-thermal instability model in WZ Sge systems, and, by
extension, in all SU UMa stars.

\subsection{Short period systems}

For systems with periods less than about one day, such as A
0620-00, the whole disc is affected by the outburst. One therefore
expect that the superhump modulation, if related to the 3:1 tidal
instability, should be visible in these systems. Systems for which
a superorbital modulation has been claimed \citep[\object{XTE
J1118+480}, \object{Nova Muscae 1991}, \object{GRO J0422+32} and
\object{GS 2000+25};][]{z02,oc96} all have periods shorter than
one day. However, the disc is always larger than the 3:1 resonance
radius as confirmed by the presence of the superhump in quiescence
\citep{z02}, hence outbursts in SXTs cannot be related to tidal
interactions. Interestingly superhumps in quiescence have been
also observed in SU UMa stars \citep{patt95}.

\subsection{Long period systems}

In systems with long ($>$1 day) orbital periods, the outer disc
will not be affected by the outburst, except possibly by
illumination effects. Hence if the tidal instability sets in, it
will always remain present, and superhumps are expected to be
present both in quiescence or in outburst. It turns out however
that in quiescence, the outer disc is too cold to radiate
efficiently even in the infrared: most of the luminosity
originates from the central regions, despite the reduced emitting
area. Figure \ref{fig:sxt} shows the structure of a disc in a
system with the orbital parameters of ``typical" long period
binary (primary mass: 12 M$_\odot$; secondary mass : 0.7
M$_\odot$; orbital period: 78 hr); the mass transfer rate was
taken to be $10^{16}$ g s$^{-1}$. We assume here $\alpha_{\rm
cold} = 0.02$ and $\alpha_{\rm hot} = 0.2$. The irradiation of the
disc is treated as in \citet{dhl01}: we assume that the disc is
illuminated by an X-ray flux that is proportional to the X-ray
luminosity; we take here {\bf $\sigma T_{\rm irr}^4 = 5\times
10^{-3} L_{\rm x}/4 \pi r^2$,} and solve for the vertical disc
structure assuming that the disc effective temperature is such
that it allows for reradiating the X-ray flux and the viscous
flux. We also assume that the inner disc is truncated by some
mechanism, presumably evaporation and/or formation of an ADAF, and
that the inner disc radius has the same functional form as the
Alfv\'en radius with a magnetic moment of 5 10$^{31}$ G cm$^3$.

In quiescence, the effective temperature is of order of 1000 K,
and less than 0.1 \% of the total luminosity is emitted by the
outer parts of the disc; any modulation of the light emitted by
these cool regions would therefore be undetectable.

In outburst, the disc heats up as a result of irradiation from the
primary (we assumed here the same prescription for irradiation as
in \citet{dlhc99} and in \citet{dhl01}). A significant fraction of
the disc luminosity in the infrared band should therefore be
emitted by regions of the disc possibly affected by the tidal
instability. However, the opacity is minimum for temperatures of
the order of 3,000 K, typical in these regions that are affected
by illumination, and it turns out that the optical depth is very
small. As a consequence, the assumption of blackbody emission does
no longer hold, and the colour temperature will be significantly
higher than the effective temperature; light will be shifted in
the optical or blue band, and will again be diluted in the total
light emitted by the inner disc.

We therefore do not expect to detect any superorbital modulation
in these systems, in the hypothesis that it would originate from
the outer parts of the accretion disc (whether it is caused by a
tidal instability or not).

\section{Conclusion}

The effects of the companion on the accretion disc are still not
well explained. Even in the simple case where no strong resonance
is present, observations of VY Scl stars indicate that the tidal
torques must be strong enough to truncate the disc at radii much
smaller than the so called tidal truncation radius. This calls for
a theoretical reexamination of the tidal torques in binaries, an
enterprize beyond the scope of this paper. Such reexamination
should also take into account on results of SPH codes. The
presence of superhumps during the superoutburst of \object{U Gem}
casts doubt on the validity of the tidal-thermal model of these
phenomena. This conclusion is strengthened by theory and
observations of superhumps in TV Col. If we make the assumption
that superhumps are in some way related to the 3:1 resonance, we
predict that superhumps should be observed in short period SXTs in
quiescence and in outburst, but not in long period systems. The
outburst mechanism should then be unrelated to this resonance, and
the tidal-thermal instability model does not apply to these
systems, casting even more doubts on the applicability of the
tidal-thermal instability to SU UMa systems.

\acknowledgements{}

We thank Michael Truss for interesting comments and information on
SPH calculations of tidally distorted discs.

\end{document}